\documentclass[traditabstract,twocolumn]{aa}
\usepackage{graphicx,amssymb}
\usepackage{txfonts}
\usepackage{natbib}
\bibpunct{(}{)}{;}{a}{}{,} 
\newenvironment{lyxlist}[1]
  {\begin{list}{}
    {\settowidth{\labelwidth}{#1}
     \setlength{\leftmargin}{\labelwidth}
     \addtolength{\leftmargin}{\labelsep}
     }}
  {\end{list}}
\usepackage[T1]{fontenc}
\newcounter{mytabs}
\begin{document}

\title{Neutron-induced  astrophysical reaction rates for translead nuclei
}

\author{I.~V. Panov \inst{1,2,3} \and I.~Yu. Korneev \inst{3} \and T. Rauscher \inst{1}
\and G. Mart\'inez-Pinedo \inst{4,5}  \and  A. Keli\'c-Heil \inst{4}
 \and N.~T. Zinner \inst{5}
   \and   F.-K. Thielemann \inst{1}
\\}
\offprints{I.V.~Panov, \email{Igor.Panov@itep.ru}}

\institute{
 Department of Physics, University of Basel, Klingelbergstr. 82,
  CH-4056 Basel, Switzerland\\
\and Institute for Theoretical and Experimental Physics,
  B. Cheremushkinskaya St. 25, 117259, Moscow, Russia\\
  \and
      Russian Research Center Kurchatov Institute, pl. Kurchatova 1, Moscow, 123182,
      Russia\\
  \and
  GSI Helmholtz Center for Heavy Ion Research, Planckstr. 1, D-64291 Darmstadt,
 Germany\\
  \and
  Department of Physics, Harvard University, Cambridge, MA 02138, USA
  }

\date{Received: / Accepted}

\abstract{ Neutron-induced reaction rates, including fission and neutron
capture, are calculated in the temperature range $10^{8} \leq T({\rm K})\leq
10^{10}$ within the framework of the statistical model
  for targets with  the  atomic number   $84\leq Z\leq 118$
(from Po to Uuo) from the neutron to the proton drip-line.  Four
sets of rates have been calculated, utilizing - where possible -
consistent nuclear data for neutron separation energies and fission
barriers from Thomas-Fermi (TF), Extended Thomas-Fermi plus
Strutinsky Integral (ETFSI), Finite-Range Droplet Model (FRDM) and
Hartree-Fock-Bogolyubov (HFB) predictions.  Tables of calculated
values as well as analytic seven parameter fits in the standard
REACLIB format are supplied$^0$. We also discuss the sensitivity of
the rates to the input, aiming at a better understanding of the
variations introduced by the nuclear input. }


\titlerunning{Neutron-induced  rates...}

\authorrunning{Panov \& et al}

\maketitle

\section{Introduction}
\label{sec:intro}
Investigations of nucleosynthesis processes make use of
reaction networks including thousands of nuclei and tens of
thousands of reactions. Most of these reactions occur far from
stability and thus cannot yet be directly studied in the laboratory.
 In addition most of the nuclear properties including reaction rates,   which are  also required
 for the calculation of cross sections and astrophysical reaction
 rates, are not experimentally known either. Therefore, predictions based on
 theoretical models are necessary. While close to stability partial
 experimental information is available, relying fully on theoretical
 information leads to relatively large
variations  in computed cross
 sections far from stability.
 This is especially true for the region of
fissionable nuclei, which is the focus of the present investigation.

      In the past, a series of
efforts were applied to calculate   neutron-capture rates for r-process
nucleosynthesis and other astrophysical applications    \citep[e.g.,][and
references therein]{A72,HWF76,WFH78,sar82,thi87,cowan91,rafkt00,most05,gori08}.

 Fission has often been neglected in astrophysical calculations.
 In early applications to astrophysical  nucleosynthesis, usually only one mode was
considered,
 beta-delayed fission \citep{TMK83} or a phenomenological model of spontaneous
  fission   \citep{gori99,frei99,cowan99}.
  However, it was shown recently that neutron-induced fission is more
important  than beta-delayed fission  in r-process nucleosynthesis
\citep{pan2002,panfkt04,fiss07}. Thus, the need to provide a compilation of
neutron-induced fission rates is obvious.   Initial investigations have
been undertaken by \cite{pan05} and \cite{talys09}. Here we
  present extended calculations of   neutron-induced fission rates for
  different
  predictions of masses and fission barriers.
            The present work also completes existing
nuclear neutron-capture  rate sets  by extending the works of \cite{rafkt00}
and \cite{pan05} to the region  $84\leq Z\leq 118$  in order to provide the
necessary input for nucleosynthesis studies under high neutron densities.
 As in \citet{pan05}, the statistical model
approach of Wolfenstein-Hauser-Feshbach \citep{Wol51,HF52} for compound nuclear
reactions was used, but employing more recent and complete data and predictions
for masses, spins, and fission barriers.
\footnotetext{Tables 3-18 with these data are only available in electronic
form at the CDS via anonymous ftp to cdsarc.u-strasbg.fr
(130.79.128.5) or via
http://cdsweb.u-strasbg.fr/cgi-bin/qcat?J/A+A/}

Nuclear mass and fission barrier predictions have a strong model dependence, and
none of the existing models can reproduce all experimentally known data.
Moreover, the fission process itself is complicated, and extended calculations
for neutron-induced fission across the nuclear chart  have to be done
carefully.
Here, we aim to provide rates for studying the endpoint of the
r-process and the possible production of super-heavy elements. By
comparing rates obtained with different  choices of mass and
fission barrier predictions we attempt to give a measure of the
involved variations. Astrophysical models, providing the
nucleosynthesis conditions, bear large  variations in
themselves. This is especially true for the r-process, for which the
astrophysical site is still unknown  despite decades of study.
For a realistic and exhaustive exploration of synthesis conditions,
simulations do not only have to vary astrophysical parameters, but
also have to include a variation range of involved reaction
rates   given by different mass and fission barrier models.

Our paper is structured as follows. In Sect. \ref{sec:statmod} we  briefly
describe
  the statistical model used in the calculations as well as the nuclear
input data and give a comparison of cross sections or rates for  a number of
experimentally known nuclei with existing experimental information and other
theoretical models.
         These methods are then applied to supplement
the rate sets of \citet{rafkt00} of $(n,\gamma)$-rates for chemical elements
with $Z>83$ and predict neutron-induced fission cross sections and rates (where
available in comparison to experiments). Section \ref{sec:rates} presents these
results and shows the sensitivity with respect to mass models and fission
barriers employed. Rate fits for utilization in astrophysical calculations are
discussed in Sect. \ref{sec:rates_2}. In Sect. \ref{sec:yields} we give a
brief discussion and some examples of the mass distribution of fission
fragments, which will be provided in an extended way in a forthcoming paper.
The final Sect. \ref{sec:conclusion} contains conclusions and a summary. The
explanation of the tables and their structure are given in Appendix {\bf A}.
  The complete tables of reaction rates and their fits are found   at CDS in
electronic form.
 \begin{figure*}
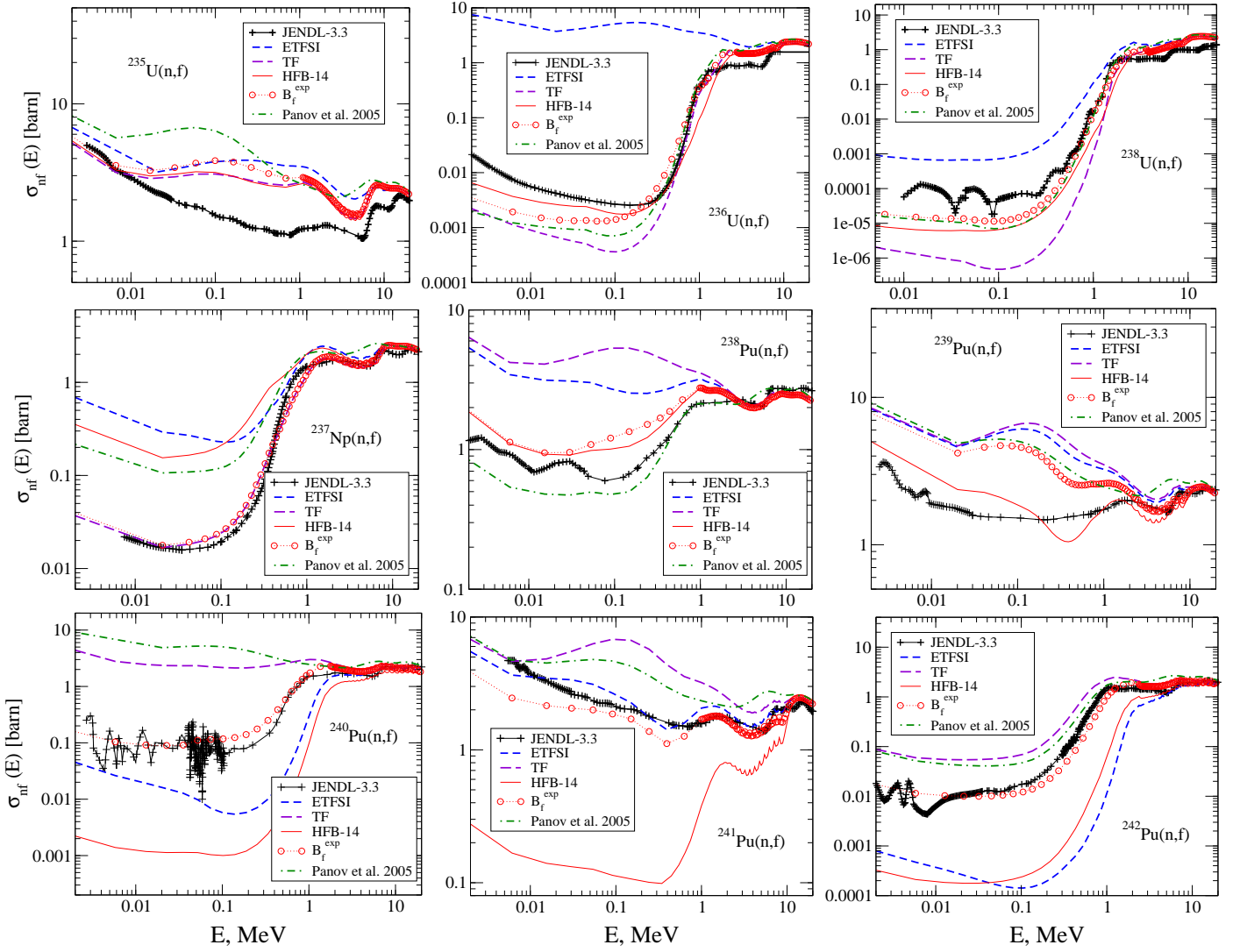

\begin{center}
  \hspace{3.5mm}\includegraphics*[width=.32\textwidth]{11967fg01a.eps}
 \hspace{0mm}\includegraphics*[width=.32\textwidth]{11967fg01b.eps}
  \hspace{0mm} \includegraphics*[width=.325\textwidth]{11967fg01c.eps}
 \hspace*{2mm}\includegraphics*[width=.335\textwidth]{11967fg01d.eps}
 \hspace{2mm}\includegraphics*[width=.305\textwidth]{11967fg01e.eps}
 \hspace{3mm} \includegraphics*[width=.305\textwidth]{11967fg01f.eps}
\hspace*{0mm}
\includegraphics*[width=.345\textwidth]{11967fg01g.eps}
\hspace{1mm}
\includegraphics*[width=.305\textwidth]{11967fg01h.eps}
  \includegraphics*[width=.325\textwidth]{11967fg01i.eps}
  \end{center}
  \caption{  Present predictions  of energy-dependent $(n,f)$ cross sections
$\sigma_{nf}(E)$ for some target nuclei  of U, Np and Pu calculated in the
framework of different mass and fission barrier predictions (ETFSI, TF, HFB-14)
and experimental data, marked $B_{\rm f}^{\rm exp}$ as well. Experimentally
measured cross-sections  were used after JENDL-3.3 \citep{jendl05}, averaged by
the code JANIS \citet{janis07}, displayed by a black
  line. All the predictions are given for a ground-state population.   Our previous
   results \citep{pan05} are shown as well. }
  \label{ex_jeja}
\end{figure*}

\section{The statistical model and nuclear data input}
\label{sec:statmod}

As in previous approaches \citep[e.g.,][]{thi87,cowan91,rafkt00}
 we have applied the
statistical Wolfenstein-Hauser-Feshbach formalism \citep{Wol51,HF52}
 for the calculation of neutron-induced cross sections
and reaction rates. In addition to $(n,\gamma)$-reactions the
fission channel was also included as outlined in \cite{TMK83,thi89},
\cite{cowan91}, \cite{pan05}. The statistical model is applicable for
astrophysical rate calculations as long as there is a sufficiently high
density of excited states in the compound nucleus at the relevant
bombarding energy, which is the case for most heavy nuclei. However, at
shell closures and with decreasing neutron separation energies, level
densities at the astrophysically relevant compound formation energy in
neutron-induced reactions become too small for the application of the
model, as shown by \citet{RTK97}. In those cases, single resonances and
contributions from the direct reaction mechanism have to be taken into
account \citep{raubi98,gorkhan}. This underlines on the one hand that
reliable mass predictions for the separation energies are absolutely
necessary for r-process applications far from stability, and on the
other hand that the prediction of spectroscopic properties for resonant
and direct capture awaits improvement as well. Here we calculate
neutron-induced rates still based purely on the statistical model
    to provide a full set of rates for extended r-process
calculations and the possible formation of superheavy nuclei for a
variety of different sets of mass models and fission barrier
predictions. The influence of the direct reaction mechanism on the
rates far from stability needs to be explored separately in a future
study.

We     outlined the general treatment within the statistical model for
applications, the cross section for a neutron-induced reaction
$i^0(n,out)$ ("out" standing for gamma-emission or fission) from the
target ground state $i^0$ with center of mass energy $E_{in}$ and
reduced mass $\mu_{in}$   given by
\begin{eqnarray}
  \sigma_\mathrm{(n,out)}^{0}=
 \frac{ \pi \hbar^2}{ (2\mu_{in}E_{in})  } \times
   \frac{1 }{  (2J^{0}_i+1)\cdot(2J_n+1)}
    \times  \nonumber
    \end{eqnarray}
\begin{equation}
\qquad  \qquad \sum_{J,\pi} (2J+1)  \frac{T^{0}_n (E,J^{\pi},
E^0_i,J^0_i,{\pi}^0_i)
   T_{out} (E,J^{\pi}) }
   { T_{\mathrm{tot}}(E,J^{\pi})}.
  \label{eq_signf}
\end{equation}
The total transmission coefficient \( T_{\mathrm{tot}}=\sum _{\nu ,o}T_{o}^{\nu
} \) describes the transmission into all possible bound and unbound states \(
\nu  \) in all energetically accessible exit channels \( o \) (including the
entrance channel $i$).  The fission transmission coefficient $T_f(E,J^{\pi})$
includes the sum over all possible final states and is evaluated    as
discussed in \cite{lynn80}, \cite{cowan91}, \cite{pan05} and  is related
to the fission probability      $P_f (E,J^{\pi})=T_f (E,J^{\pi})/
T_{\mathrm{tot}}(E,J^{\pi})$   considered in the papers cited above. Since the
work of \citet{strut} fission has been generally described within the framework
of double-humped fission barriers.  When making use of a double-humped fission
barrier, the fission transmission coefficients can be calculated in the limit
of complete damping, which averages over transmission resonances, assuming that
levels in the second minimum between the first and second barrier $E_A$ and
$E_B$ are equally spaced. If detailed information concerning the level
structure in the second well is missing, this method gives the best results.
$P_f$ further requires the transmission coefficients through the first and
second barrier $T_A$ and $T_B$, which are evaluated by an integral over the
first and second barrier potential, weighted with the level density at the
appropriate energy and corresponding deformation. The individual barriers can
be approximated by individual (inverted parabola) Hill-Wheeler barrier shapes.
 The level densities $\rho_A$ and $\rho_B$ show an enhancement over the level
densities at ground state deformation, and it is important to include proper symmetry classes in  the
calculations  at corresponding saddle points.
   In the absence of detailed information, standard factors of  four  (for the axially
asymmetric/mass symmetric barrier) and two (axially symmetric/mass asymmetric)
over the ground state level density  were  applied \citep{lynn80}.
 Whenever possible, experimentally known
fission barriers were used, taken from \citet{IAEA_smir},   the
  compilation of \citet{mamdo98}, and  the database  of \cite{ripl2}. The other transmission
coefficients were calculated as in \citet{rafkt00}, utilizing up to 19
experimentally known excited states (if available). The data  were  taken
from \citet{toi96}, up to the first level for which the spin assignment
is   unknown. Ground state spins and parities are known for many
unstable nuclei. Far off stability, ground state spins and parities  were
taken from \citet{mnk97} when experimental values  were unavailable.
Above the last known state, the nuclear level density of \citet{RTK97}
  was  used.    
      This method is based on the back shifted Fermi-gas approach, where the level
density parameter $a$ and back shift $\delta$ are obtained globally
from the appropriate mass model employed.

   In Fig.\ \ref{ex_jeja} we compare our predictions for
neutron-induced fission cross sections of some U, Np and Pu isotopes
with evaluated neutron data from JENDL-3.3  \citep{jendl05,janis07}.
   (The accuracy of the evaluated data  is  usually not declared. In the region
of interest, $0.01<T_9<10$ MeV, the accuracy of up-to-date
measurements for plutonium isotopes  by \cite{exper09pu}  varies  from
2\% to 15\%, and for our plots is not bigger than  the plot signs.  The
detailed experimental information can be  found in the  \cite{exfor2009}).
Experimental masses and fission barriers or masses and
fission barriers were employed  from different mass models:  
  ETFSI \citep{etfsi,mamdo98}, TF \citep{mysw96,mysw99}, HFB-14
\citep{talys09} and the older liquid drop predictions by
\citet{homo80}, here shown as Panov et al. 2005). It can be seen
that   when using experimental fission barriers, the agreement with
experimental $(n,f)$ cross sections is within the typical factor of
  two to three known for statistical model calculations. It can also be seen
that different barrier predictions can lead to large
variations, which will clearly remain for predictions far from
stability where no experimental information is available. These
results can also be compared to a recent investigation by the
  Reference Input Parameter Library (RIPL)
community \citep{talys09}, making use of the code TALYS and HFB-14
fission barriers plus nuclear level densities obtained from a
combinatorial approach based on single particle spectra from the
corresponding Hartree-Fock-Bogolyubov calculations.  The fission
barriers in \citep{talys09} were   employed   following the fission
path via a    Wentzel-Kramers-Brillouin   phase integral,
which should be superior to a Hill-Wheeler
inverted parabola treatment. When examining their Fig.\ 8, which is
based on pure predictions, it can be recognized that our results
with HFB-14 fission barriers
    employed via a double-humped fission barrier approach    are similar or
even closer to the  experiment than the results given therein. As we make use of the
same barriers, we relate the difference mainly to the different level density
predictions (here a back-shifted Fermi gas with parameters obtained from a
global mass model, there a combinatorial approach to single particle spectra
from a microscopic HFB mass model). We draw the same conclusions from the
comparison with their Fig.\ 9 and our calculations with experimental barriers.
While microscopic investigations  should be   in principle more advanced, the
back-shifted Fermi gas approach based on global mass models seems still more
robust in its predictive power. While renormalizations of fission paths and
level densities can improve this picture (see their Figs. 10-12), this is only
possible if experimental data are available. As our investigations are meant
for astrophysical applications far from stability, where no experimental
information is available, we come to the conclusion that our approach is well
suited for this endeavor. It  does, however, depend  on the quality of mass and
fission barrier predictions,  and this will be the focus of the present work.

In astrophysical plasmas, reactions also occur on thermally excited
target states. The stellar cross section $\sigma^*$ can then be defined
as a sum of  the cross sections $\sigma^x$ for those excited states $x$
with excitation energy $E_x$ and spin $J_x$, weighted by the Boltzmann
excitation probability
\begin{equation}
\label{eq:stellarcs}
 \sigma^*=\frac{\sum_x {\left( 2J_x+1 \right) \sigma^x e^{-\frac{E_x}{kT}}}}
 {\sum_x {\left(2J_x+1 \right) e^{-\frac{E_x}{kT}}}} \quad .
\end{equation}

 \begin{figure*}
\begin{center}
 \includegraphics*[width=.45\textwidth]{11967fg02a.eps}
 \hspace{1.cm} \includegraphics*[width=.45\textwidth]{11967fg02b.eps}
  \end{center}
  \caption {Comparison of $(n,\gamma)$-rates (integrated over
    Maxwell-Boltzmann distributions of targets and projectiles for the
    displayed temperatures) from present calculations   and
    other existing predictions for the target nuclei $^{238}$U (left) and $^{242}$Pu
    (right) with experiment.   The symbols are chosen as follows:
    \citet{most05} (blue dash-dot line), \citet{gori08} (thin blue line),
     JENDL-3.3 \citep{jendl05} (crosses),   new
    Talys-based  \citep{talys09}   predictions (green dash-dot  line).   Our
    calculations are shown for reaction rates including only the
    ground state (GS) or a thermally populated target (GS+tps). Only
    the GS rates can be compared to experimental data. The results of
    \citet{most05,gori08} correspond to GS + tps conditions.}
  \label{U_Pu_sigv_ng}
\end{figure*}

 \begin{figure*}
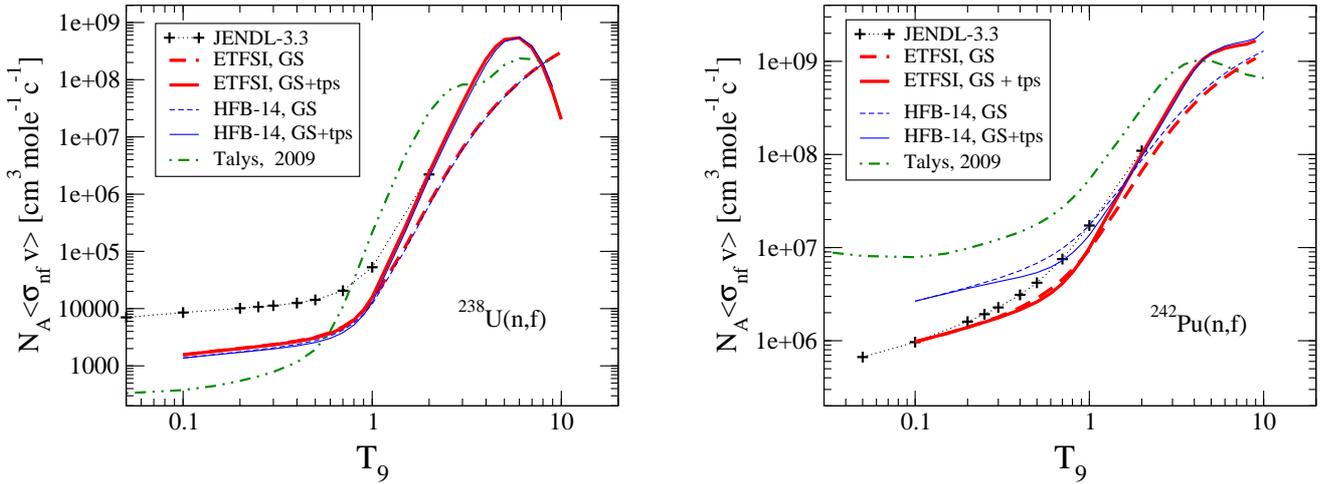

\begin{center}
 \includegraphics*[width=.44\textwidth]{11967fg03a.eps}
 \hspace{1.cm}  \includegraphics*[width=.44\textwidth]{11967fg03b.eps}
  \end{center}
  \caption {  Comparison of $(n,f)$-rates (integrated over
    Maxwell-Boltzmann distributions of targets and projectiles for the
    displayed temperatures) from present calculations   and
    other existing predictions for the target nuclei $^{238}$U (left) and $^{242}$Pu
    (right) with  the experiment. The symbols are chosen as follows:
    \citet{talys09} (green two-dot-dash line),
     JENDL-3.3 \citep{jendl05} (crosses).  Our
    calculations are shown for reaction rates including only the
    ground state (GS) or a thermally populated target (GS+tps) for different nuclear data
    predictions (HFB or ETFSI).   The results of
    \citet{talys09} correspond to GS + tps conditions.}
  \label{U_Pu_sigv_nf}
\end{figure*}

The $\sigma^x$  were  calculated in the same way as shown in Eq.\
(\ref{eq_signf}) for the ground state, i.e.\ for $x=0$. Only the
stellar cross sections can be used to compute the appropriate
astrophysical reaction rates. The reaction rate for a specific
reaction at a given stellar temperature \( T \)  was  then determined
by folding the stellar reaction cross section \( \sigma \)\(
^{*}(E) \) with a Maxwell-Boltzmann distribution of relative
velocities between projectiles and targets \citep{fowler}:
\begin{equation}
\label{eq_rate}
 \left\langle \sigma ^{*}v\right\rangle =\left\langle
\sigma v\right\rangle ^{*}=
      \left( \frac{8}{\pi \mu }\right) ^{1/2}\frac{1}{\left( kT\right)
^{3/2}}\int\limits _{0}^{\infty }\sigma ^{*}(E)E\exp \left( -\frac{E}{kT}
\right) dE  .            
\end{equation}

    Figures \ref{U_Pu_sigv_ng}  and  \ref{U_Pu_sigv_nf} show   a
typical comparison of  present neutron capture
  rate calculations of $^{238}$U and $^{242}$Pu to experimentally based rates from
  JENDL-3.3
\citep{jendl05} and the predictions of \citet{most05,gori08,talys09}. The
agreement between rate predictions and data for other nuclei in this mass range
are of the same order. Note that these predictions  along the valley of
stability are based on experimental masses, which leads to an average agreement
with experimental cross sections within a factor of 1.5. The expected
variation will be larger far from stability where theoretical mass and barrier
predictions have to be utilized.

\section{Neutron-induced fission rates for a variety of mass models}
\label{sec:rates}

Early r-process
calculations \citep{TMK83}, which included fission, made use of the
mass predictions by \citet{hilf76}
  and the fission barriers of a macroscopic-microscopic
model by \citet{homo80}. For many years
different authors used the fission barriers from \citet{homo80},  as they were
the only complete set of barriers available. More recently, renewed
interest (and increased computing power) spurred a number of new
calculations of large sets of barrier predictions within various
models, resulting on "average" in higher values of fission barriers
than predicted by \citet{homo80}. For a
consistent treatment of nucleosynthesis, fission rates
should be calculated with the neutron separation energies,
reaction Q-values and fission barrier heights derived from the
same mass model \citep[see the discussion in][]{cowan91,rauapp94}.

 \begin{figure*}[t!]
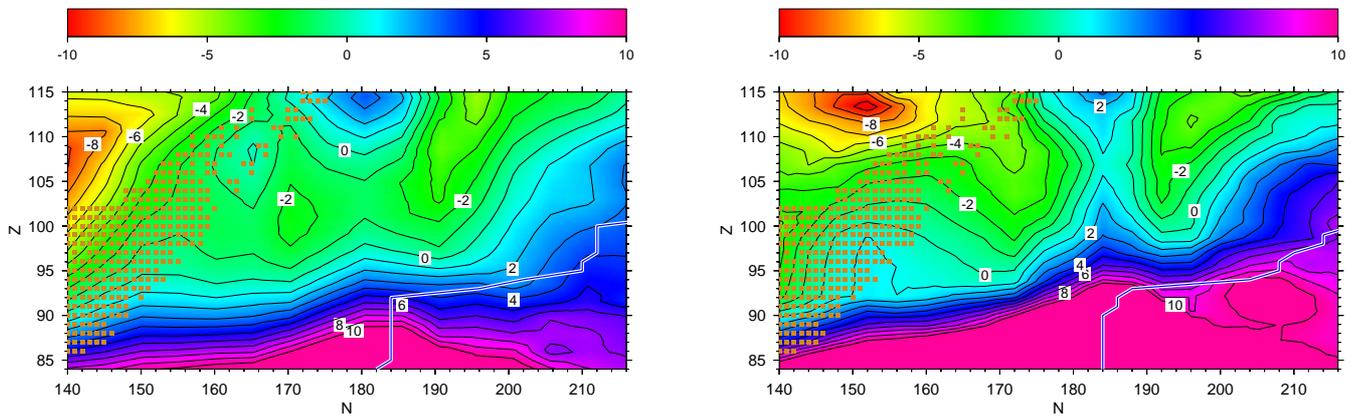

\begin{center}
 \includegraphics*[width=.45\textwidth]{11967fg04a.eps}
 \hspace{1.cm}  \includegraphics*[width=.45\textwidth]{11967fg04b.eps}
  \end{center}
  \caption{  Contour plot of the quantity $B_f-S_n$ (in MeV) for two choices of models
for masses and fission   barriers: FRDM masses plus TF fission barriers (left)
and ETFSI-predictions (right). Also indicated are known superheavy nuclei
(boxes) and the drip-line (line), where $B_f-S_n$ determines whether the
r-process continues towards heavier nuclei or fission cycling to lighter
fission products occurs.}
   \label{barriers}
\end{figure*}

  As explained in Sect. \ref{sec:statmod}, different mass- and fission
barrier predictions were utilized to test the sensitivity stemming from
different underlying models. The models used are: masses taken from the Finite
Range Droplet Model (FRDM) by \citet{frdm}, the Extended Thomas-Fermi with
Strutinsky Integral (ETFSI) model by \citet{etfsi}, and the Thomas-Fermi (TF)
model of \citet{mysw96}; fission barriers are taken from the ETFSI
\citep{mamdo01} and TF \citep{mysw99} models. It should be emphasized that the
ETFSI masses  employed here
(http://www-astro.ulb.ac.be/Nucdata/Masses/etfsi2-plain) are based on the force
 SkSC18  \citep{ref-sksc18},
 while the
 ETFSI fission barriers (http://www-astro.ulb.ac.be/Nucdata/Fisbar/fisbar1)
 were obtained with the force SkSC4 \citep{mamdo98}.
 Thus, it is difficult
to perform fully consistent calculations, and the quality of barrier heights   is not really known
especially in the region far from experimentally known nuclei, where  the
r-process proceeds.
 For this reason the choice of two
different sets of fission barriers, TF and the ETFSI, permits to test the
sensitivity range. When utilized together with mass predictions from the same
models (see   however   the remarks in the previous paragraph), a reasonably
consistent treatment is possible. In addition, we also chose to explore a
combination of TF fission barriers  and FRDM nuclear mass predictions and
to gauge the resulting effect. The latter are close to the TF predictions,
containing also the same shell corrections \citep{mysw96} and have already been
used in some astrophysical rate calculations for nuclei with $Z<84$.

Figure \ref{barriers} shows the quantity of primary importance, $B_f-S_n$, for
the calculation of neutron-induced fission cross sections, based on the two
model sets FRDM+TF and ETFSI. $B_f-S_n$ indicates the  regions of nuclei
where neutron induced fission rates can be high enough ($B_f-S_n<0$) and  are
important for  the r-process nucleosynthesis (see  also the discussion of Fig.\ 6).
We see that both sets display a quite different behavior    and note that
$B_f-S_n$ is generally larger for ETFSI than FRDM+TF.

 \begin{figure*}
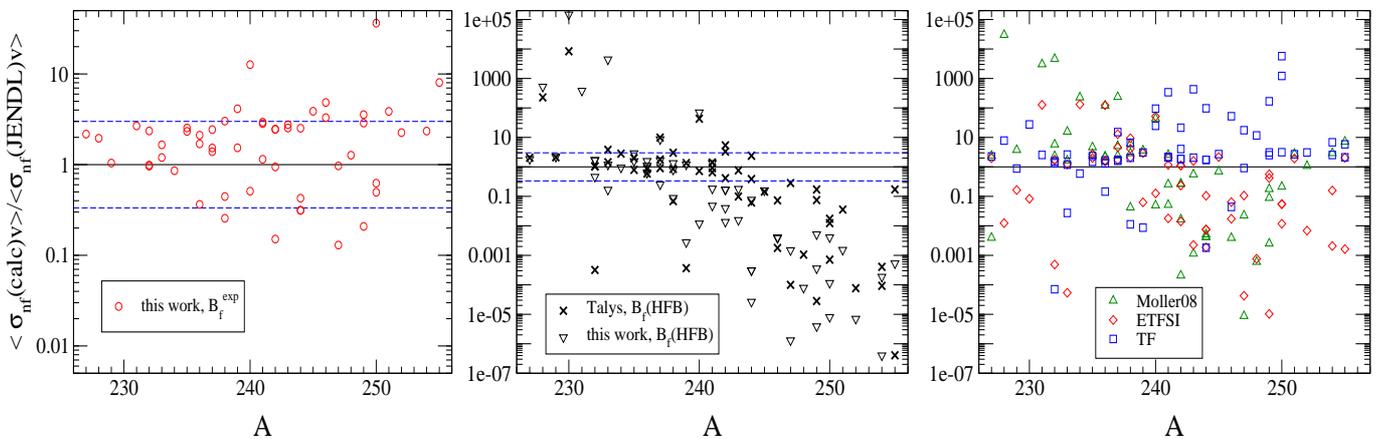

\begin{center}
 \includegraphics*[height=.31\textwidth,width=5.9cm]{11967fg05a.eps}
 \hspace{0.1cm}\includegraphics*[height=.31\textwidth,width=5.9cm]{11967fg05b.eps}
 \hspace{0.1cm}  \includegraphics*[height=.31\textwidth,width=5.9cm]{11967fg05c.eps}
  \end{center}
  \caption{    Calculated maxwellian averaged neutron-induced fission cross sections
   in comparison to evaluated experimental  fission cross sections at
    30 keV.                 
 Left panel: experimental fission barriers were utilized in
    the calculations (circles) of maxwellian averaged cross-sections in comparison to experimental
    values  for 58 isotopes from Th to Fm    taken from JENDL-3.3 \citep{jendl05};
 central panel:
the same ratio of calculated   and evaluated
    experimental   values, but employing theoretical fission
    barrier predictions from different sources:  the Extended Thomas-Fermi
    model by \citet{mamdo01} (diamonds),  the  Thomas-Fermi by \citet{mysw99}
    (squares),  and  recent predictions from \citet{Moller08}
    (green triangles up).
    When not experimentally known, consistent nuclear
    masses were utilized from the corresponding  model predictions.
  Right panel: comparison
   of maxwellian averaged cross-sections from our calculations (triangles)
     and from  the  Talys \citep{talys09}       calculations  (crosses), both utilizing
      HFB predictions    (with BSk14 Skyrme force)  of masses and fission
      barriers.
    }
   \label{stat_exp}
\end{figure*}

Our fission cross section calculations were performed within
the framework of a double-hump fission barrier (permitting the
existence of a double or single hump barrier). The
heights of both barriers f were  predicted in the calculations of
\citet{homo80} (HM) and \citet{mamdo01}. However, the TF model predicts
only one of the fission barriers. In order to employ
  this model in our calculations we  assumed  that the predicted
  fission barrier is the higher of both barriers and
  derived the lower one as described below and in \citet{pan05}.
In order to derive the height of the lower barrier, we compared two
methods: (i) keeping the difference in height of the first and second barrier
of the original HM barriers and (ii) keeping the same height ratio, i.e.\
relative height, of the barriers. The resulting cross sections were not
very different because the heights of the lower barriers calculated in these
two ways differ only by a few  percent for the majority of cases. Only in a
small number of cases the cross sections differ by more than 10\% (but the
largest deviations do not exceed a factor of two). For the rates shown here we
chose to use the difference in height of the first and second barrier of HM to
predict the lower barrier for the TF approach. In this manner, the new fission
rate calculations were extended to the region above charge Z=100, whereas
\citet{homo80} presented results only for $Z\leq 100$. This allows us to
perform r-process calculations in extremely neutron-rich environments as well
as to study  superheavy element production in rapid neutron capture
nucleosynthesis processes.

An extended comparison to evaluated neutron-induced fission cross sections,
based on evaluated data \citep{jendl05} for the trans-lead region is shown in
 Fig.\ \ref{stat_exp}.  The left panel displays the ratio of predicted to
 evaluated cross section when using experimentally known fission barriers
(circles).     The agreement of  the  calculated rates based on the experimental
values of fission barriers is quite good with the majority of ratios,   and is
within factors of  two to three. Some outliers of up to a factor of ten  are observed in a
few cases, but the experimental accuracy of the barrier determination is not
known for these cases.        The middle panel  shows  the difference between  the
calculated ones   when  the  HFB mass and fission barrier predictions were used: in
this work (triangles down) and by Talys (crosses).  Existing  difference in two
Hauser-Feshbach calculations  can  have  emerged from  the  differences in fission barriers
values (due to renormalization by \citet{talys09}), different level density
used etc.  In the right panel of Fig.\ \ref{stat_exp}  the calculated cross
sections made use of mass and fission barrier predictions from TF and ETFSI    as
well as new predictions by \citet{Moller08}.

 Contrary to the comparison with the
left panel (circles), when using only experimental barriers we  found  that
the difference between calculated cross sections and
     measured ones can be many orders of magnitude (up to a factor $10^4$).
It is interesting to note that a weak systematic can be seen. Cross sections
calculated with ETFSI as well as HFB fission barrier predictions show a large
scatter above and below the measured values for the lighter end of the plotted
mass range, whereas they tend to underpredict the fission cross sections for
heavy masses.   The calculations using the recent barriers by \citet{Moller08}
 seem to show a similar behavior with a smaller scatter, but they are currently
available only for a more limited number of nuclei. The results obtained with
the TF barriers exhibit a different pattern, the predicted cross sections  agree
well
 for the lighter mass nuclei, but seem to be systematically too
large for the heavier ones.

\begin{table*}   
\caption{Neutron separation energy $S_n$ and fission barrier predictions for
nucleus $^{262}$U formed after neutron capture by $ ^{261}$U. }
 {\begin{tabular}{|c|rrr|rrr|rrc|rrc|rrc| }
\hline\hline
 Mass and  $B_f$  models& \multicolumn{3}{|c|}{ETFSI} & \multicolumn{3}{|c|}{TF}  &
\multicolumn{2}{|c}{TF} &  {FRDM} &
\multicolumn{2}{|c}{TF} &  {ETFSI}& \multicolumn{3}{|c|}{HM }  \\
\hline  $^{262}U$ & $B_{\rm f1}$  & $B_{\rm f2}$ & $S_{\rm n}$
 & $B_{\rm f1}$ & $B_{\rm f2}$ & $S_{\rm n}$
 & $B_{\rm f1}$ & $B_{\rm f2}$   & $S_{\rm n}$ & $B_{\rm f1}$ & $B_{\rm f2}$& $S_{\rm n}$ & $B_{\rm f1}$
 & $B_{\rm f2}$   & $S_{\rm n}$  \\
\hline\hline
 {\rm calculated  values,  in MeV} & 3.9 &  5.30 & 4.46 &  1.20 & 4.56 & 4.05    &  1.20 & 4.56 &
 3.81  &  1.20 & 4.56 &  4.46
   & 0 &  3.36 & 4.14\\
\hline\hline
\end{tabular}}
\label{tabl_mf}
\end{table*}

The above comparison underlines the  considerable  variations still inherent in
fission barrier predictions. However, we suggest that by comparing TF with
ETFSI \citep[and with][]{Moller08} predictions, the relevant variation
range can be estimated. Theoretical cross sections depend strongly on the
  fission barriers, and a high accuracy for their values is required. Because
of the impact of the fission barrier uncertainties, nucleosynthesis
studies at present should explore a variety of barrier sets, while
waiting for further measurements and improved predictions. For this
reason  we compute and compare  below  rates for different
sets of nuclear properties and also provide tables and fits of the
rates for all these cases.

\begin{figure*}
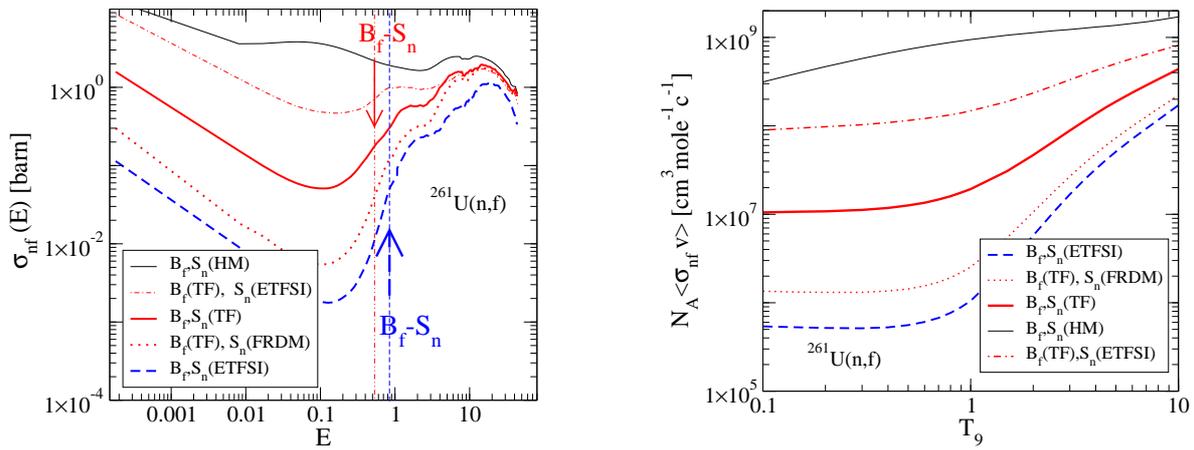

\begin{center}
 \hspace*{-0.5cm} \includegraphics*[width=.38\textwidth]{11967fg06a.eps}
 \hspace{1.5cm}  \includegraphics*[width=.38\textwidth]{11967fg06b.eps}
  \end{center}
  \caption{Dependence of neutron-induced fission cross sections $\sigma_{nf}(E)$ (left)  and rates
  $\lambda_{nf} = N_A \left\langle \sigma v \right\rangle $(right) on mass-
  and fission barrier predictions
  for $^{261}$U. The sources for fission barriers $B_f$ and neutron separation
energies $S_n$ (Howard and M\"oller, Thomas-Fermi, extended
Thomas-Fermi, finite range droplet model)
are indicated in the panel. Arrows show the difference
between fission barrier and neutron separation energy $B_f-S_n$ for
ETFSI (dashed line) and TF (full line)
  predictions.
 }
   \label{u_261e_mf}
\end{figure*}

Figure \ref{u_261e_mf} shows fission cross sections (left panel) and
rates (right panel) calculated by combining different sets for predictions
of masses and fission barriers. The arrows in the left plot show the
difference between the fission barrier height and the neutron
separation energy $B_f-S_n$ given by predictions of the TF model (red
arrow at the top of the left panel) and the ETFSI model (dashed arrow
at the bottom). The exact values can be found in Table \ref{tabl_mf}.
The cross sections (and thus the rates) depend essentially on the
available energy $B_f-S_n$ in the fission channel, minor dependencies
on fission barriers heights and neutron separation energies, individually,
are due to the competition with the $(n,\gamma)$-channel.

It can be seen from Fig.\ \ref{u_261e_mf} that the neutron-induced fission
cross sections (as well as the rates) increase with decreasing
fission barriers and that $\sigma_\mathrm{nf}^\mathrm{HM} >
\sigma_\mathrm{nf}^\mathrm{TF} > \sigma_\mathrm{nf}^\mathrm{ETFSI}$.
The difference at low energies is due to the different mass
predictions used (comparing calculations with TF fission barriers
but different neutron separation energies $S_\mathrm{n}$). The small
decrease in
 $S_\mathrm{n}$,  when predictions for masses and fission barriers (based on
 the two sets of input data) are changed from TF+TF to FRDM+TF,
  results in a decrease of the neutron-induced fission cross sections
 (accompanied by decreasing $(n,\gamma)$-cross sections). The same
 influence is illustrated by the cases  where  sets of consistent
determinations for $S_n$  and $B_f$ are replaced by sets from different mass
predictions.
 For example, different predictions of neutron separation energies
 were used  for the same TF -- fission barrier predictions.
 Therefore, cross sections for low energies differ significantly,
   but for higher $ E  $ the difference in cross section becomes much smaller for the
   same fission barriers (here we used TF barriers).
   The temperature averaged rates (Fig.\ \ref{u_261e_mf}, right panel) show the same
 dependence.

\begin{figure*}
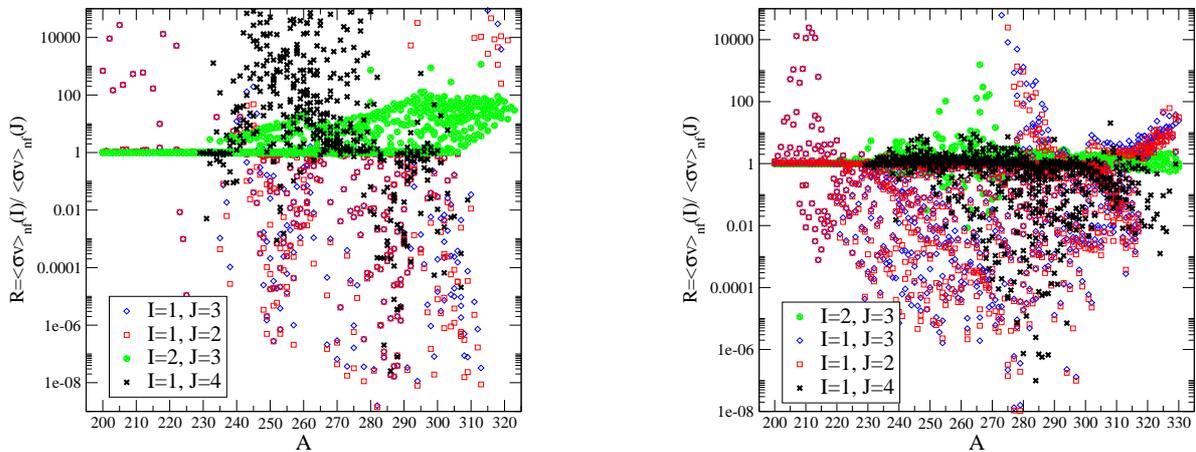

\begin{center}
 \includegraphics*[width=.368\textwidth]{11967fg07a.eps}
 \hspace{2cm}  \includegraphics*[width=.368\textwidth]{11967fg07b.eps}
  \end{center}
  \caption{ Ratio $R=\frac{<\sigma v>_I}{<\sigma v>_J}$ of reaction rates
for different fission barrier and mass predictions at temperatures
 $T_9=0.3$  (left) and $T_9=10 $  (right). The
indexes I,J define the mass/barrier predictions utilized in the
calculations:
    ETFSI    (I,J=1), Thomas-Fermi (I,J=2) or Thomas-Fermi for fission-barriers and FRDM for
  mass-predictions (I,J=3), Hartree-Fock-Bogolyubov (HFB-14) predictions
  (I,J=4).
  As the Thomas-Fermi model and the Finite Range
Droplet Model give similar mass predictions, the deviation for the combination
2/3 is the smallest (and also decreases with temperature). However,  the
use of   TF (or FRDM) vs ETFSI (or HFB-14) predictions changes the rates
drastically
 and thus the full range of possible magnitudes  shown in Fig.\ \ref{barriers}  
 is obtained.
 }
 \label{r_stat4}
\end{figure*}

As expected, the rate sets calculated on the basis of FRDM masses and the TF
model for masses and fission barriers are quite comparable (Fig.\
\ref{r_stat4}). On the other hand, when the sets include ETFSI vs. FRDM+TF or
TF comparisons, the results can differ  by  up to  eight  orders of
magnitude. The most extreme difference of ETFSI-based and other rates was
obtained for nuclei with neutron numbers close to 184, for which the ETFSI
model predicts very high fission barriers, leading to small fission rates.  The
difference between rates, calculated on the basis of ETFSI and HFB mass and
fission barriers predictions is less than between ETFSI and TF, especially for
higher $T_9$. For smaller $T_9$ the we can see that difference of rate values
has the opposite sign for regions with A $\approx  240-280$ and $A>280$. This
 agrees  with Fig.\ \ref{stat_exp}.        Some of the combinations
employed in Fig.\ \ref{r_stat4} are     shown to underline the huge problems
which can arise when not using consistent data.  The extrapolation of rate
calculations to regions of very exotic nuclei is a hard task, and only further
investigations can answer which kind of prediction is more preferable. At the
moment the only choice is to test all available predictions in r-process
calculations  that are  compared to astronomical abundance observations.

\begin{table*}
 \caption{ Ranges of isotopes for which we calculated rates based on input
from the models TF, FRDM, ETFSI and HFB. Given are the charge number \protect\(
Z\protect \) and the lower and upper limits \protect\( A_{\mathrm{min}}\protect
\) and \protect\( A_{\mathrm{max}}\protect \) of the neutron number of the
targets in the isotopic chain.}

\begin{tabular}{|c|cc|cc|cc|cc||c|cc|cc|cc|}
\hline\hline
   & \multicolumn{2}{|c|}{TF-shc}        & \multicolumn{2}{|c|}{FRDM}       & \multicolumn{2}{|c|}{ETFSI}
&  \multicolumn{2}{|c||}{ HFB} &
 &\multicolumn{2}{l}{TF-shc}          & \multicolumn{2}{|c|}{FRDM} & \multicolumn{2}{|c|}{ETFSI} \\
\hline $Z$ & $A_{\rm min}$ & $A_{\rm max}$ & $A_{\rm min}$ & $A_{\rm
max}$ & $A_{\rm min}$ & $A_{\rm max}$ & $A_{\rm min}$ & $A_{\rm max}$
&  $Z$&
  $A_{\rm min}$ & $A_{\rm max}$ &$A_{\rm min}$ & $A_{\rm max}$ & $A_{\rm min}$ & $A_{\rm max}$ \\
\hline\hline   
 84$^*$ &  180 & 269 & 180 & 269 & 180 & 267 &193 & 269  &102 & 233 &  331 &  233 & 331 & 233&  331\\
 85$^*$ &  185 & 270 & 185 & 270 & 185 & 270 &193 & 270  &103 & 240 &  335 &  240 & 334 & 240&  328 \\
 86$^*$ &  187 & 269 & 187 & 269 & 187 & 269 &196 & 269  &104 & 239 &  337 &  239 & 337 & 239&  324 \\
 87$^*$ &  190 & 280 & 190 & 280 & 190 & 280 &200 & 278  &105 & 242 &  337 &  242& 331 & 242&  330\\
 88$^*$ &  193 & 283 & 193 & 283 & 193 & 279 &202 & 279  &106 & 245 &  337 &  245 & 331 & 247&  337\\
 89$^*$ &  196 & 288 & 196 & 288 & 196 & 284 &206 & 285  &107 & 248 &  337 &  248 & 332 & 243&  337 \\
 90 &  199 & 293 & 199 & 293 & 199 & 287 &220 & 287 &108 & 251 &  337 &  251 & 327 & 245&  337 \\ 
 91 &  212 & 296 & 212 & 296 & 212 & 288 &229 & 289 &109 & 254 &  337 &  254 & 327 & 247&  337\\
 92 &  204 & 299 & 204 & 299 & 204 & 291 &230 & 291 &110 & 257 &  337 &  257 & 339 & 249&  337\\
 93 &  220 & 302 & 220 & 302 & 220 & 293 &232 & 293 &111 & 260 &  337 &  260 &331 & 251&  329\\
 94 &  210 & 305 & 210 & 305 & 210 & 295 &234 & 295 &112 & 263 &  337 &  263 &332 & 253&  329 \\
 95 &  215 & 309 & 215 & 308 & 215 & 297 &238 & 297 &113 & 267 &  337 &  267 &327 & 255&  329 \\
 96 &  216 & 312 & 216 & 311 & 216 & 300 &240 & 300 &114 & 270 &  337 &  270 &327 & 257&  329\\
 97 &  225 & 315 & 225 & 314 & 225 & 303 &243 & 315 &115 & 273 &  337 &  273 &337 & 275&  329\\
98 &  222 &  319 &  222 & 319 & 222& 317 &245 & 319 &11 6& 276  &  337 &  276 &337 &  - &   - \\
99 &  225 &  322 &  225 & 322 & 225& 320 &250 & 322 &117 & 279  & 337 &  279 &337 &  - &   -  \\
100&  227 &  325 &  227 & 325 & 227& 325 &251 & 325 &118 & 282  &  337 &  283 &337 &  - &   - \\
101&  239 &  328 &  239 & 328 & 239& 326 &254 & 328 &119 &  -   &   -  &   -  & -  &  - &   - \\
 \hline\hline
\end{tabular}
\\ $^*$ { for these chemical elements  there are no   HFB-predictions of fission barriers }
\label{rangeZA}
\end{table*}

\section{Computed rate sets and mass ranges}
\label{sec:rates_2}

As we discussed in the introduction, our aim was to prepare sets of
neutron-induced reaction rates, suited for calculations of the r-process up to
very high atomic masses.  In this sense, our rates extend the previously
published ones for   $Z<84$ \citep{rafkt01} up to the region  $84\leq Z
\leq Z_{max}$  ($Z_{max}$ depends on available nuclear input and varies from
102
 for HFB-predictions  to 118 for FRDM data).  The exact
range of nuclei for the different choices of mass and fission barrier
input is given in Table \ref{rangeZA}.

Our calculations include all outgoing channels and give simultaneously
predictions for neutron-induced fission, (n,$\gamma$)-, (n,p)-, and
(n,$\alpha$) rates. However, here we only provide tables and fits for the
neutron-induced fission and neutron capture rates. Due to their small values
for neutron-rich nuclei, the rates of the other calculated reaction types are
not important in astrophysical applications.

  The format of the tables is explained in Appendix A. The full electronic
versions of the tables available on-line  at the CDS include all  rates for all
mass predictions (Tables  3--6). The isotope   and element ranges for which
rates as well as   rate fits are available are given in Table \ref{rangeZA} for
the FRDM, TF, ETFSI and HFB mass predictions.  This amounts to 2151 (ETFSI),
2637 (TF),  2400 (FRDM-masses, TF-barriers)   and 1323 (HFB) involved nuclei.
The partition functions for all isotopes are given on a grid of 24
temperatures: \( T_{9} \) = 0.1, 0.15, 0.2, 0.3, 0.4, 0.5, 0.6, 0.7, 0.8, 0.9,
1.0, 1.5, 2.0, 2.5, 3.0, 3.5, 4.0, 4.5, 5.0, 6.0, 7.0, 8.0, 9.0, 10.0 and can
be found in CDS's Tables 3--6 as well.  We also provide the fit coefficient
needed to compute the photodisintegration rate (see Sect.\ \ref{sec:fits}  ).

\subsection{  The fits for neutron-induced and reverse  rates }
\label{sec:fits}

Reaction rates have been calculated  on  a grid of 24 temperatures:
$T_9$=0.1, 0.15, 0.2, 0.3, 0.4, 0.5, 0.6, 0.7, 0.8, 0.9, 1.0, 1.5, 2.0, 2.5,
3.0, 3.5, 4.0, 4.5, 5.0, 6.0, 7.0, 8.0, 9.0, 10.0, same as for partition
functions. These rates include the thermal modification in the stellar plasma,
i.e.\ consider reactions from excited states in the target. For easy
application in astrophysical investigations, these stellar rates were fitted
with the same REACLIB parameterization as used for other reaction types earlier
\citep{rafkt00}:
\begin{eqnarray}
N_{A}\left\langle \sigma v\right\rangle
    ^{*}
=\exp \left( a_{0}+a_{1}T_{9}^{-1}+a_{2}T_{9}^{-1/3}
+a_{3}T_{9}^{1/3}+a_{4}T_{9}     \right.  \nonumber  \\
\left. +a_{5}T_{9}^{5/3}+a_{6}\ln T_{9}\right) \quad , \label{eq_fitpar}
\end{eqnarray}
with the seven open parameters \( a_{0}-a_{6} \) and the stellar
temperature \( T_{9} \) given in 10\( ^{9} \) K. This parameterization
  proved  to be flexible enough to accommodate the different temperature
dependencies of the various reaction types across the fitted
temperature range of \( 0.01\leq T_{9}\leq 10. \) Parameterizations of
the present rates in the form used in \citet{HWF76} can be found in
Appendices.
The best fit  was  obtained by minimizing the
 deviation $\zeta$ (Eq.\ (\ref{eq_dev})) using the FUMILI code \citep{silin}.

The flexibility of the fitting function makes it prone to numerical
problems outside the calculated range at low temperatures. In some
cases they tend to diverge strongly. This difficulty can be avoided by
 additionally providing fit data at low temperatures  to the calculated
values by appropriately extrapolating the rates to lower temperatures.
However, it has to be emphasized that  the considered parameterization
is  only valid within the temperature range of \( 0.01\leq T_{9}\leq
10.0 \), although many fits will show a ``proper'' behavior down to
lower temperature.

\begin{figure*}
 \hspace*{1.5cm}\begin{center}
 \includegraphics*[height=6.5cm]{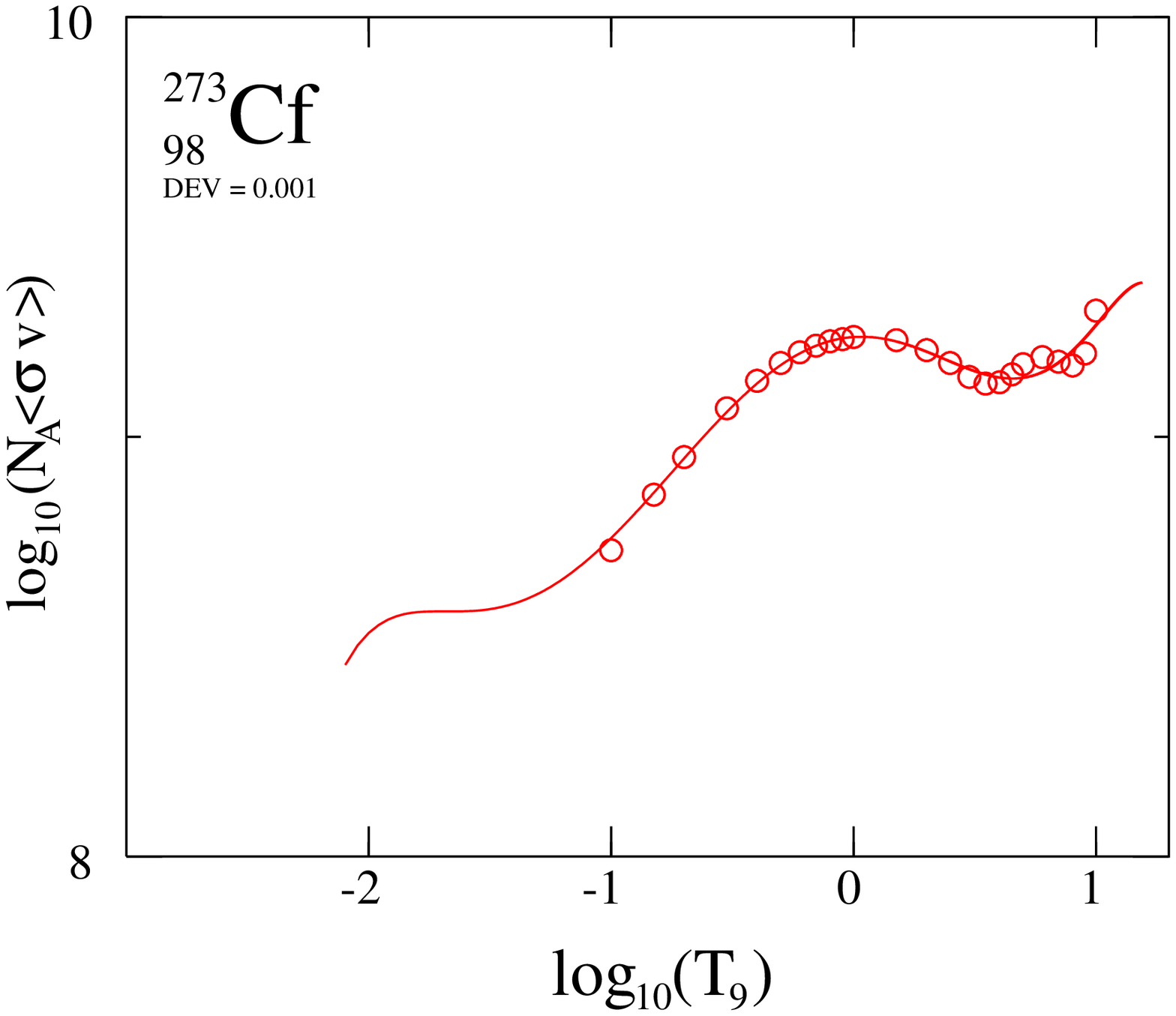}
 \hspace{2cm}\includegraphics*[height=6.5cm]{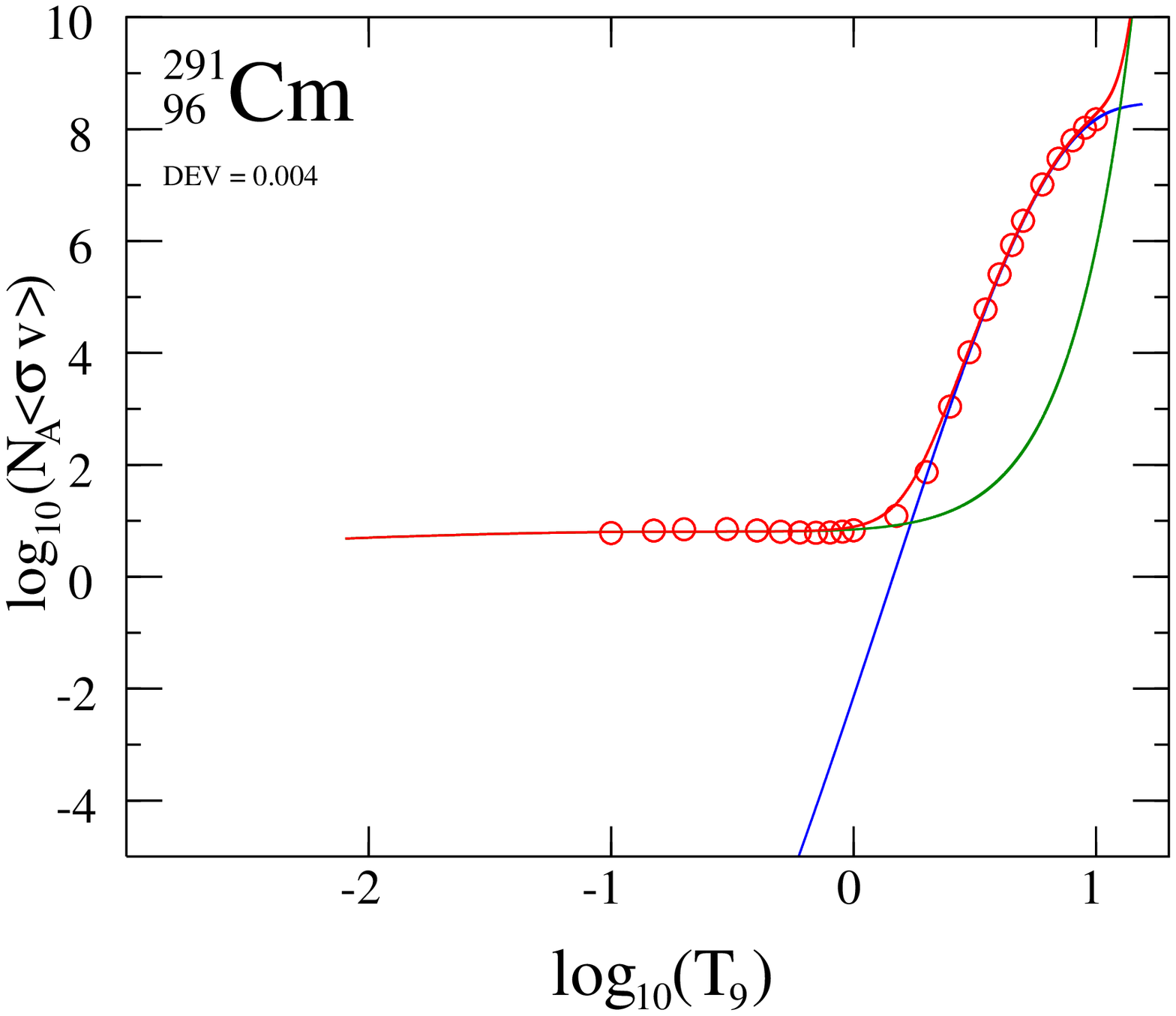}
  \end{center}
  \caption{Representative fits of neutron-induced fission rates ($^{273}$Cf
 and $^{291}$Cm) for two types of temperature  dependences. The general aim
is to attain sufficient accuracy with one set of seven fitting
coefficients (see Eq.\  (4)), as shown in the left curve. In some cases
the superposition of two such sets is necessary (right panel)
 to correctly reproduce the low- as well as high-temperature
behavior. A correct extrapolation to low temperatures
($0.01<T_9<0.1$) is important to avoid unphysical abundance changes
      (see also the explanation of tables).
 }
  \label{fits1}
\end{figure*}

\begin{figure*}
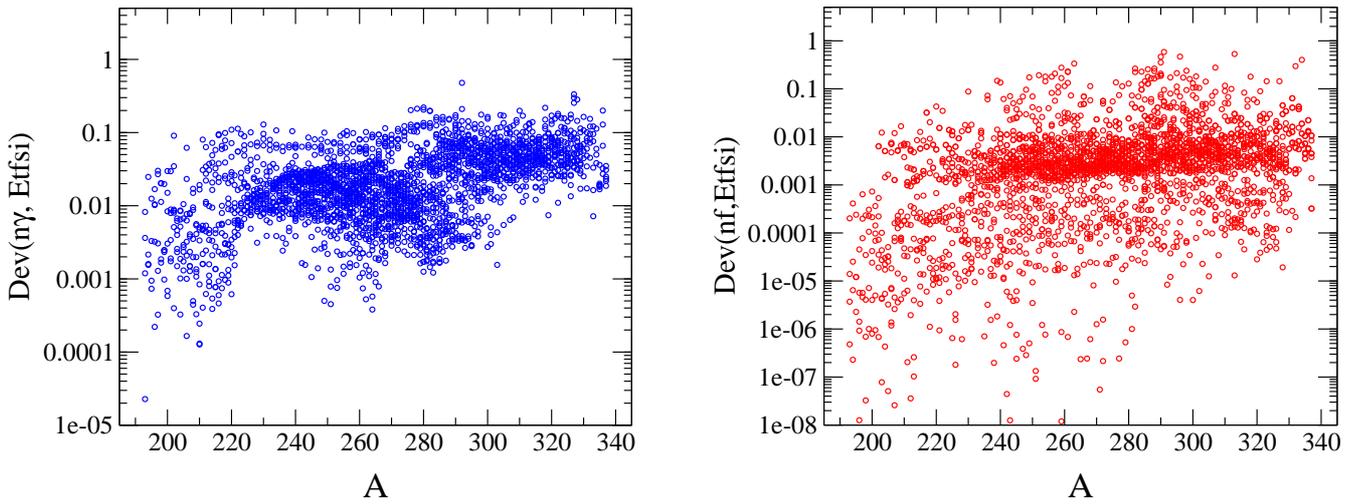

\begin{center}
 \includegraphics*[width=.45\textwidth]{11967fg09a.eps}
 \hspace{1cm}\includegraphics*[width=.45\textwidth]{11967fg09b.eps}
  \end{center}
  \caption{The accuracy of fits is defined by a mean-square error
(see Eq.\ (\ref{eq:accur})), shown here for
  $(n,\gamma)$-rate fits (left panel) and neutron-induced fission rate fits
(right panel). We see that the upper boundary of errors is in the
range of 10\%. This should be compared to the variations of
 these  calculations (see Figs.\ \ref{ex_jeja} and \ref{stat_exp}).
 }
   \label{fits_accur}
\end{figure*}

As a measure of the accuracy of a given fit, the quantity \( \zeta  \)
(marked in tables and figures as {\it Dev}) is used. It is
defined by
\begin{equation}
\label{eq:accur} \zeta =\frac{1}{24}\sum _{i=1}^{24}\left(
\frac{r_{i}-f_{i}}{f_{i}}\right) ^{2}\quad , \label{eq_dev}
\end{equation}
with \( r \) being the original rate value as calculated at each of
the 24 temperatures $T_9$= 0.1, 0.15 \dots  10.0, and   $f_i$   is
the rate calculated from the fit at these temperatures. A small
value of $\zeta$ indicates an accurate fit over the entire
temperature range.  Higher  values of $\zeta$ are mainly caused by
deviations at low temperatures, where rates are slow and a larger
deviation is permissible. For the majority of nuclei the value of
$\zeta$ is less than 1 and lies in the range $0.1-10^{-4}$ (see
Fig.\ \ref{fits_accur}).   We should mention here that the
modified approximation formula by  \cite{cyburt08}  can probably   give
the better fit.
   Its accuracy can be high  and should be applied for reactions,
    whose  rates can be calculated with rather high accuracy.
    For our case fitting  the  average accuracy is not bad (see Fig.\ \ref{fits1}),
    and is   much better than accuracy of
    calculations of neutron-induced rates for very neutron rich nuclei.
     We applied the approximation Eq.\ (\ref{eq_fitpar}),
    used earlier in a number of previous predictions of neutron rates.

The temperature dependence of the rate can be one of two
types, as illustrated in Fig.\ \ref{fits1}.         
         These types of behavior can be understood  when one recalls
the discussion of Fig.\ \ref{u_261e_mf}. A fission transmission
coefficient which is constant or slowly varying as a function of energy
leads to an $(n,f)$-cross section which (similar to a pure neutron
capture) shows a $1/\sqrt{E}$ dependence, if s-wave dominated.
Averaging such a cross section over a Maxwell-Boltzmann distribution
yields a constant rate.  This situation occurs  for
example  when the neutron bombarding energy leads to a compound nucleus
energy above the lower and below the higher barrier of a double-hump
fission barrier. Then the lower barrier is open and the penetration
through the remaining higher barrier is close to constant, but the
height of the higher barrier determines the size of the $(n,f)$ cross
section.  This  behavior is seen in Fig.\ \ref{u_261e_mf} below about 0.1 MeV (left
panel). In the right panel the corresponding rate is
shown and seen to be close to constant below about $T_9=1$.

In contrast, the opening of a fission barrier as a function of energy,
i.e. an exponentially increasing transmission coefficient close to the
barrier energy, leads to a sudden rise of the cross section, and
consequently also of the rate as a function of temperature. We see this
behavior for bombarding energies  on the order of  $B_f-S_n$ in the left
panel of Fig.\ \ref{u_261e_mf} and a similar behavior for the rates in the right panel.
The size of this change from an almost constant to a steeply rising
rate is a function of the barrier height. The right panel of Fig.\ \ref{fits1} is a
very representative example of such a case (almost constant rate at low
temperatures and a steep rise by orders of magnitude at a critical
temperature). On the other hand, when the compound  energy is
close to  the barrier height already for
small bombarding energies, the cross
section and rate are already large at small energies (temperatures).
Such an example can  be seen in the left panel of Fig.\ \ref{fits1}, reflecting a
double-hump  behavior with small energy differences between the lower
and higher barrier.
       For  a discontinuity in the $T_9$-dependence (right panel), the
fit was performed as a sum of two contributions and is given by two lines in
the   Table \ref{Tab_a2} of the paper and CDS's Tables   7--18.

For all cases it is recommended to use the fits only down to the
temperature $T_9\geq 0.01$. Moreover, close to the drip-line, the
statistical model may not be applicable for reactions with low $Q$
value, even above that temperature. Although the fit may be good, the
user should be aware of that possible complication.

\subsection{Photodisintegration rates and partition functions}
\label{sec:revpar}

For a full implementation of the neutron captures in a reaction network, the
inverse photodisintegration rates also have to be known. The
photodisintegration rates (and their fits) are not explicitly given in the
tables, but can be computed from the information contained therein. To calculate
the reverse rate of the reaction  B(n,$\gamma$)D, i.e.\ the reaction
D($\gamma$,n)B, the seven parameters
$a_{0}^{\mathrm{rev}}-a_{6}^{\mathrm{rev}}$ are determined as follows:

\begin{eqnarray}
  a_{0}^{\mathrm{rev}} & = &  a_0+ ln \left(9.8685 \times
10^{9}
  \left(\frac{A_\mathrm{D}-1}{A_\mathrm{D}} \right)^{1.5}
\frac{2J_\mathrm{B}+1}{2 J_\mathrm{D}+1}  (2 J_\mathrm{n}+1)   \right)      \quad,    \nonumber\\
a_{1}^{\mathrm{rev}} & = & a_{1}-11.6045S_n  \quad,  \nonumber\\
a_{2}^{\mathrm{rev}} & = & a_{2} \quad, \nonumber\\
a_{3}^{\mathrm{rev}} & = & a_{3} \quad, \nonumber\\
a_{4}^{\mathrm{rev}} & = & a_{4} \quad, \label{revcoff}\\
a_{5}^{\mathrm{rev}} & = & a_{5} \quad, \nonumber\\
a_{6}^{\mathrm{rev}} & = & a_{6}+1.5 \quad,
 \nonumber
\end{eqnarray}
with $A_\mathrm{D}$    the mass number of nucleus D, $J_\mathrm{n}$
     the   spin of neutron and $J_\mathrm{B},J_\mathrm{D}$      the
ground state spins of nuclei B and D, respectively. These parameters
for the reverse reaction are also given in the tables (see explanations
in the appendices).

It is important to note that the value computed by applying Eq.\
(\ref{eq_fitpar}) with the above coefficients has to be multiplied by the
ratio of the partition functions for  the  residual and target nucleus \(
G_\mathrm{B}/G_\mathrm{D} \) to obtain the actual photodisintegration
rate. Examples are shown in
Appendix \ref{sec:examples}. As it was shown in detail earlier
\citep{rafkt00}, the temperature-dependent partition function \(
G(T^{*}) \) normalized to the ground state spin \( J^{0} \) of a nucleus
is defined as in \citet{fowler67}
\begin{eqnarray}
(2J^{0}+1)G(T^{*})= &
 \sum _{\mu =0}^{\mu _{m}}(2J^{\mu }+1)
e^{-E^{\mu }/kT^{*}}     \\
 & +\int\limits _{E^{\mu _{m}}}^{E^{\mathrm{max}}}\sum _{J^{\mu }, \pi ^{\mu
}}(2J^{\mu }+1) e^{-\epsilon/kT^{*}}\rho (\epsilon ,J^{\mu },\pi ^{\mu
})d\epsilon \quad ,  \nonumber
\end{eqnarray}
with $\rho$ being the level density and \( \mu  \)\( _{m} \) the last included
experimentally known state. For the temperature range considered here, the
maximum energy \( E_{i}^{\mathrm{max}} \) above which there are no more
significant contributions to the partition function is  on  the order of \( 20-30
\) MeV \citep{rafkt00}. The temperature dependent partition functions  are
available  at the CDS, as well as the fit coefficients in Eq.\ (\ref{eq_fitpar})
(Tables 7-18) for the $(n,\gamma)$, $(\gamma,n)$ and (n,f) rates.

This subsection discussed photo-induced reaction rates as inverse
reactions of
neutron capture rates, which are presented in full detail
in the present publication. Photo-induced reactions can also lead to fission
when the Planck distribution of photons provides a significant fraction of
photons with energies above the fission barrier. This reaction channel is
not discussed here but its possible influence in r-process environments
will be analyzed in a future investigation.

\section{Fission fragment distributions}
\label{sec:yields}

A proper inclusion of fission in r-process calculations also
requires the knowledge of the resulting distribution of fission
fragments, which have to be entered as reaction products. This by
itself requires a major effort and will be presented with a
thorough description of the treatment plus detailed fission yield
distributions in a forthcoming paper. However, at the end of the
present investigation we want to give a short outlook on how this
topic will be approached.

In principle, such distributions are dependent on the excitation energy of the
compound nucleus   and thus would lead to a changing yield distribution for each
bombarding energy. However, we found that the distributions  vary
smoothly and slowly as a function of the excitation energy. Thus, for the
neutron energy range in astrophysical applications, the yield distribution at
the neutron separation energy (i.e., for a vanishing neutron bombarding energy)
is a very good approximation. This would permit  us to multiply the
fitted (n,f)-rates from the previous section with a static distribution of
yields for all temperatures.

Here we only briefly show a few examples of such fission distributions.
Properties of fission fragments, i.e., masses, atomic numbers,
excitation and kinetic energies, were calculated based on the
macro-microscopic approach (similar to the FRDM model) and the
separability of compound-nucleus and fragment properties on the
fission path \citep{schmidt08,wilkins76}. The original technical
description of the fragment-formation model was published in
\citet{benlliure98} and \citet{kruglov02}. In the calculations shown
here we used an updated description that will be the subject of a
forthcoming publication.

In the model it is assumed that the different ways of splitting up the total
mass are basically determined by the number of available transition states
above the potential energy surface behind the outer saddle point. The
macroscopic properties of the potential-energy landscape of the fissioning
system are attributed to the strongly deformed fissioning system, which are
deduced from mass distributions at high excitation energies \citep{rusanov97}
and Langevin calculations \citep{nadtochy02}. The microscopic properties of the
potential-energy landscape of the fissioning system are given by the
qualitative features of the shell structure in the nascent fragments. They are
determined from the observed features of the fission channels \citep{brosa90}
according to the procedure described by \citet{schmidt08}. The dynamics of the
fission process responsible for the fragment formation was considered in an
approximate way: It was assumed that the phase space near the outer saddle
point determines the mass asymmetry of the system, which is more or less frozen
during the descent to scission. On the other hand, it was also assumed that the
$N/Z$ collective degree of freedom is determined near the scission point. The
excitation energies of the created fragments were calculated from the available
excitation energy at the scission point and the deformation energies of the
fragments at scission. The deformation energies of the fragments were assumed
to be specific to the individual fission channels. They were deduced from
experimental data  \citep[and references therein]{wahl88, expfiss2008}
on total kinetic energies and neutron yields. Kinetic energies were then
calculated applying the energy conservation law. Finally, the two excited
fission fragments are subject to particle (mostly neutrons) and $\gamma$ ray
emission until they reach their ground state configurations. The de-excitation
process was described in the framework of the statistical model as described in
\citet{ABLA}.

\begin{figure*}
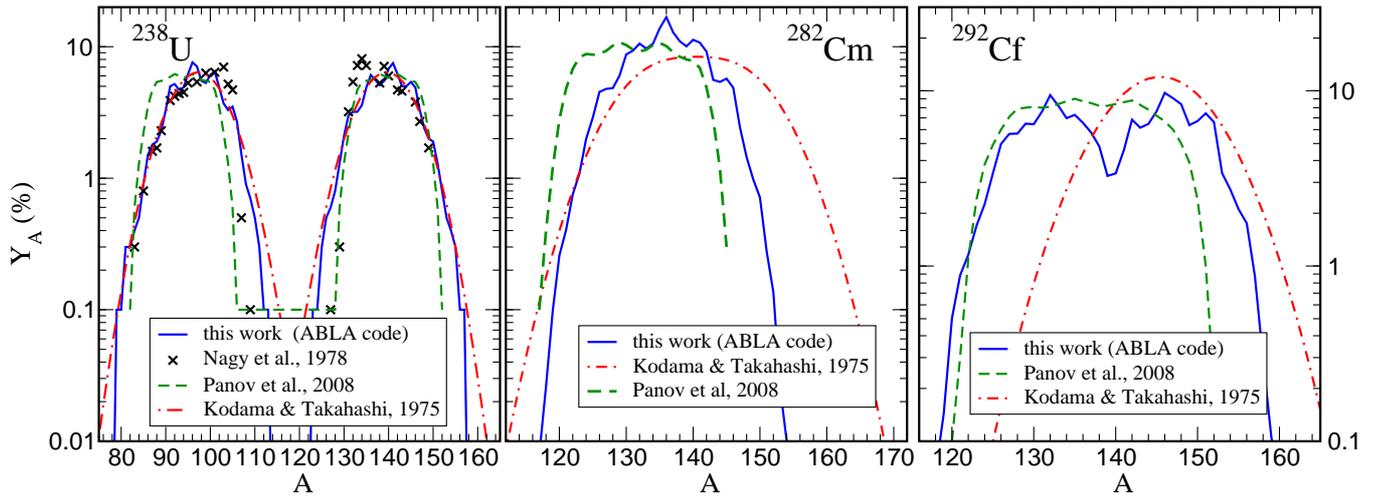
   
  \centering
  \includegraphics*[height=6.5cm]{11967fg10a.eps}
\includegraphics*[height=6.5cm]{11967fg10b.eps}
\includegraphics*[height=6.5cm]{11967fg10c.eps}
  \caption{  The  final mass distributions of fission fragments for
    the compound nuclei (after neutron capture) $^{238}$U,  $^{282}$Cm  and $^{292}$Cf.
 The distributions  were  computed  with
    the ABLA code \citep{ABLA} as described in the text. In addition, we show the
    yields computed with the phenomenological parameterizations of \citet{pan_fragm08}
    and \citet{Kodama.Takahashi:1975}, as well as experimental data (crosses).}
  \label{uyields}
\end{figure*}

 The left panel of Fig.\ \ref{uyields} compares the mass distributions
resulting from such an approach with the experimental data for $^{238}$U
\citep{exp_u238}. We also show the yields computed using the
  empirical parameterizations developed
by \citet{pan_fragm08} and \citet{Kodama.Takahashi:1975}, which previously
  were  used in r-process calculations. In order to give an impression of the
impact of the results for r-process nuclei, the two right panels of
Fig.\ \ref{uyields}  show  the mass distributions resulting from the fission of
$^{282}$Cm and $^{292}$Cf. For these nuclei, clear differences appear between
the phenomenological approaches \citep{pan_fragm08,Kodama.Takahashi:1975} and
the macroscopic-microscopic approach described above.  For application in
r-process simulations the $(n,f)$-reaction rates (which are  the focus of the
present paper) will have to be multiplied by the displayed percentages for
obtaining a production rate of a specific fragment nucleus
\citep{kelic05,fiss07,pan_fragm08}.

\section{Summary and conclusions}
\label{sec:conclusion}

We provide predictions of neutron-induced fission rates and
$(n,\gamma)$-rates for a wide range of astrophysical temperatures
($10^{8} \leq T({\rm K})\leq 10^{10}$) and targets (proton- to
neutron- drip-line for  $84 \leq Z \leq 118$, i.e. from Po to Uuo )
in the framework of the Wolfenstein-Hauser-Feshbach model, making
use of a variety of different mass and fission barrier predictions
\citep{mysw99,mamdo01,etfsi,frdm,homo80,talys09}. The astrophysical
(stellar) reaction rates were fitted   as in previous works
\citep{thi87,rafkt00}  in the common REACLIB seven parameter form,
and these parameters are also tabulated. This provides the basis for
r-process nucleosynthesis calculations where the abundance
predictions for the highest mass numbers as well as the effect of
fission cycling are strongly dependent on the interplay of neutron
capture and fission.

  In order to give an impression of the reliability of the results, we
compared them  with experiment and with available independent predictions before
exploring the currently unreachable regions of the nuclear chart with a variety
of theoretical predictions for nuclear masses and fission barriers (FRDM,
ETFSI, TF, HFB).
                 An extended comparison of
   neutron-induced fission rates with experiment
and with available independent predictions was done. The  dependence
 of rates  on
 nuclear input data, most of all fission barriers, is   high.
  Astrophysical nucleosynthesis yield predictions, especially in the
transuranium region, should  take into account these large
differences in order to explore the variations  involved. For
this reason   extended tables for neutron-induced fission rates as
well neutron capture rates are presented for different mass and
fission barrier predictions in  fitted form for nucleosynthesis
calculations. Their structure is given in the Appendix {\bf A} (note
that the full rate and fit tables are available at the CDS).
          Given that  fission predictions far from stability have not been
          tested yet, and even close to stability none of the existing models
          has yet been proven to be superior (see Fig. \ref{U_Pu_sigv_nf}).   Nucleosynthesis
          calculations should probably continue to use a variety of these models.
A further requirement for   nucleosynthesis modeling in the region of
fissioning nuclei is the  knowledge  of the   mass distribution of fission
products. This work is in progress (see Sect. \ref{sec:yields}).

\section*{Acknowledgements}

The authors thank E. Kolbe, K.-L. Kratz, K. Langanke,
 P. M\"oller, D. K. Nadyozhin, and B. Pfeiffer for useful
discussions. This work was supported by the Swiss National Science Foundation
  (SCOPES projects No.~IB7320-110996 and No.~IZ73Z0-128180/1)
and    grant 2000-105328.    
 I.P. and I.K. were partly supported by Federal Programm "Scientific and pedagogical specialists of
innovation Russia.", contract number 02.740.11.0250 and RFBR-grant
09-02-12168-ofi.

\begin{appendix} 
\label{appendix-main}

\section{Explanation of the tables and examples of how to use them }
\label{sec:examples}

\begin{table*}[t!]
\caption{   Example of format of rates presented in the   Tables  {\bf  3--6 },
available at the CDS: reaction rates  $N_{A}\left\langle \sigma v\right\rangle$
and partition functions show a subset of calculation based on ETFSI
predictions.}
   \begin{tabular}{|cc|cc|cc|cc| }
  \hline \hline
 \multicolumn{2}{|c|}{mother(A,Z)}&\multicolumn{1}{c}{$T_9$}&\multicolumn{1}{c}{p.f.}&\multicolumn{1}{|c}{(n,f)}&\multicolumn{1}{c}{(n,f)*}&\multicolumn{1}{|c}{(n,g)}&\multicolumn{1}{c|}{(n,g)*}\\
\hline\hline \label{Tab_a1}
 Cf & 273 &     0.10  &   1.00D+00 & 5.07D+08& 5.90D+08 & 2.89D+04 &3.32D+04     \\
 Cf & 273 &     0.15  &   1.01D+00 & 6.59D+08& 8.13D+08 & 3.71D+04 &4.50D+04     \\
 Cf & 273 &     0.20  &   1.02D+00 & 7.81D+08& 1.01D+09 & 4.35D+04 &5.50D+04     \\
 Cf & 273 &     0.30  &   1.08D+00 & 9.70D+08& 1.33D+09 & 5.28D+04 &6.99D+04     \\
  \ldots & \ldots &   \ldots  &   \ldots  & \ldots & \ldots & \ldots & \ldots  \\[5pt]
 Cf & 273 &     7.00  &    7.86D+04  & 1.89D+09 & 1.57D+09 &  5.70D+04 & 9.86D+01   \\
 Cf & 273 &     8.00  &    9.13D+05  & 1.93D+09 & 1.53D+09 &  5.41D+04 & 1.47D+01   \\
 Cf & 273 &     9.00  &    1.04D+07  & 1.98D+09 & 1.60D+09 &  5.14D+04 & 2.20D+00   \\
 Cf & 273 &    10.00  &    1.03D+08  & 2.03D+09 & 1.97D+09 &  4.89D+04 & 3.69D-01   \\
  \ldots & \ldots &   \ldots  &   \ldots  & \ldots & \ldots & \ldots & \ldots  \\[5pt]
 Cf & 274 &     7.00  &   3.87E+05  & 1.75E+08  & 1.68E+09  & 4.04E+06 & 5.85E+02  \\
\hline \hline
 \end{tabular}
 \end{table*}
\begin{table*}[t!]
\caption{ Parameterization of the  ($n,\gamma$)-, $(\gamma,n)$-
    and (n,f)-rates, available at the CDS Tables {\bf  6--18.}  }
\begin{tabular}{|c|c|rrrrrrr|c|}
\multicolumn{10}{c}{Example of Table {\bf  7 } --- $(n,\gamma)$-rate fits on
the basis of ETFSI mass-model predictions.
 }\\[2pt]
\hline\hline
\multicolumn{1}{|c|}{mother(A,Z)}&\multicolumn{1}{c|}{$i_{fit}$}&\multicolumn{1}{c}{$a_0$}
&\multicolumn{1}{c}{$a_1$}&\multicolumn{1}{c}{$a_2$}
&\multicolumn{1}{c}{$a_3$}&\multicolumn{1}{c}{$a_4$}&\multicolumn{1}{c}{$a_5$}
&\multicolumn{1}{c}{$a_6$}
&\multicolumn{1}{|c|}{Dev}\\[2pt]
\hline \hline \label{Tab_a2}
cf  268  98 & 0  &  -1.45E+01 &   0.00E+00 &  -2.94E+00 &   3.00E+01 &  -4.69E+00 &   1.40E-01 &  -6.04E+00 &    3.9E-02\\
cf  269  98 & 0  &   2.05E+01 &   1.43E-02 &  -6.95E-01 &  -3.73E+00 &   1.60E-01 &  -3.13E-01 &   7.64E-01 &    5.0E-02\\
cf  270  98 & 0  &  -2.03E+01 &   0.00E+00 &  -3.86E+00 &   3.56E+01 &  -5.02E+00 &   1.90E-01 &  -7.77E+00 &    1.9E-02\\
cf  271  98 & 0  &  -6.15E+00 &   1.68E-02 &  -3.20E+00 &   1.94E+01 &  -2.75E+00 &  -3.24E-03 &  -4.49E+00 &    1.1E-02\\
cf  272  98 & 0  &  -1.69E+01 &   0.00E+00 &  -4.68E+00 &   4.14E+01 &  -5.82E+00 &   1.64E-01 &  -9.39E+00 &    3.6E-02\\
cf  273  98 & 0  &   2.62E+00 &   1.71E-02 &  -2.67E+00 &   1.32E+01 &  -1.60E+00 &  -1.64E-01 &  -3.27E+00 &    2.2E-02\\
cf  274  98 & 0  &  -1.87E+01 &   0.00E+00 &  -5.35E+00 &   4.34E+01 &  -4.94E+00 &   2.04E-02 &  -1.03E+01 &    4.9E-02\\
cf  275  98 & 0  &   2.43E+00 &   0.00E+00 &  -1.74E+00 &   1.61E+01 &  -2.19E+00 &  -1.24E-01 &  -3.45E+00 &    1.9E-02\\
cf  276  98 & 0  &  -1.50E+01 &   0.00E+00 &  -4.72E+00 &   3.79E+01 &  -4.15E+00 &  -3.04E-02 &  -9.12E+00 &    3.8E-02\\
cf  277  98 & 0  &  -5.42E+00 &   0.00E+00 &  -3.49E+00 &   2.64E+01 &  -2.77E+00 &  -1.12E-01 &  -6.64E+00 &    3.1E-02\\
\hline\hline
\end{tabular}

\vspace{0.2cm}
\begin{tabular}{|c|c|rrrrrrr|}
\multicolumn{9}{c}{Example of Table  {\bf 11 } --- reverse $(\gamma,n)$-rate
fits    on the basis of ETFSI mass-model predictions.}
\\[2pt]
\hline \hline
\multicolumn{1}{|c|}{mother(A,Z)}&\multicolumn{1}{c|}{$i_{fit}$}&\multicolumn{1}{c}{$a_0^{rev}$}
&\multicolumn{1}{c}{$a_1^{rev}$}&\multicolumn{1}{c}{$a_2^{rev}$}
&\multicolumn{1}{c}{$a_3^{rev}$}&\multicolumn{1}{c}{$a_4^{rev}$}&\multicolumn{1}{c}{$a_5^{rev}$}
&\multicolumn{1}{c|}{$a_6^{rev}$}\\[2pt]
\hline \hline
cf  269  98 & 0  &   1.05E+01 &  -3.46E+01 &  -2.94E+00 &   3.00E+01 &  -4.69E+00 &   1.40E-01 &  -4.54E+00 \\
cf  270  98 & 0  &   4.28E+01 &  -5.82E+01 &  -6.95E-01 &  -3.73E+00 &   1.60E-01 &  -3.13E-01 &   2.26E+00 \\
cf  271  98 & 0  &   4.72E+00 &  -4.43E+01 &  -3.86E+00 &   3.56E+01 &  -5.02E+00 &   1.90E-01 &  -6.27E+00 \\
cf  272  98 & 0  &   1.61E+01 &  -5.62E+01 &  -3.20E+00 &   1.94E+01 &  -2.75E+00 &  -3.24E-03 &  -2.99E+00 \\
cf  273  98 & 0  &   8.14E+00 &  -3.73E+01 &  -4.68E+00 &   4.14E+01 &  -5.82E+00 &   1.64E-01 &  -7.89E+00 \\
cf  274  98 & 0  &   2.49E+01 &  -5.28E+01 &  -2.67E+00 &   1.32E+01 &  -1.60E+00 &  -1.64E-01 &  -1.77E+00 \\
cf  275  98 & 0  &   6.30E+00 &  -3.57E+01 &  -5.35E+00 &   4.34E+01 &  -4.94E+00 &   2.04E-02 &  -8.87E+00 \\
cf  276  98 & 0  &   2.47E+01 &  -5.01E+01 &  -1.74E+00 &   1.61E+01 &  -2.19E+00 &  -1.24E-01 &  -1.95E+00 \\
cf  277  98 & 0  &   1.00E+01 &  -3.66E+01 &  -4.72E+00 &   3.79E+01 &  -4.15E+00 &  -3.04E-02 &  -7.62E+00 \\
cf  278  98 & 0  &   1.68E+01 &  -5.02E+01 &  -3.49E+00 &   2.64E+01 &  -2.77E+00 &  -1.12E-01 &  -5.14E+00 \\
\hline\hline
\end{tabular}

\vspace{0.2cm}
\begin{tabular}{|c|c|rrrrrrr|c|}
\multicolumn{10}{c}{Example of Table  {\bf 15} ---
  neutron-induced fission rate fits on the
basis of ETFSI  mass-model predictions.
 }\\[2pt]
\hline\hline
\multicolumn{1}{|c|}{mother(A,Z)}&\multicolumn{1}{c|}{$i_{fit}$}&\multicolumn{1}{c}{$a_0$}
&\multicolumn{1}{c}{$a_1$}&\multicolumn{1}{c}{$a_2$}
&\multicolumn{1}{c}{$a_3$}&\multicolumn{1}{c}{$a_4$}&\multicolumn{1}{c}{$a_5$}
&\multicolumn{1}{c}{$a_6$}      
&\multicolumn{1}{|c|}{Dev}\\[2pt]
\hline \hline
cf  268   98 & 0 &  3.13E+01 & -6.04E-02 &  8.24E+00 & -1.81E+01 &  2.22E-01 &  7.09E-02 &  8.65E+00 & 5.2E-03\\
cf  269   98 & 0 &  3.38E+01 &  0.00E+00 &  2.22E+00 & -1.59E+01 &  1.39E+00 & -7.64E-02 &  4.52E+00 & 4.5E-03\\
cf  270   98 & 0 &  9.75E+00 &  3.15E-03 & -1.49E+00 &  1.34E+01 & -1.77E+00 &  1.59E-01 & -2.78E+00 & 7.7E-03\\
cf  271   98 & 0 &  3.84E+01 & -5.73E-02 &  8.81E+00 & -2.75E+01 &  1.86E+00 & -9.37E-02 &  1.04E+01 & 1.5E-03\\
cf  272   98 & 0 & -1.30E+01 &  0.00E+00 & -4.24E+00 &  4.22E+01 & -5.66E+00 &  4.90E-01 & -9.00E+00 & 1.7E-02\\
cf  273   98 & 0 &  3.87E+01 & -5.66E-02 &  8.78E+00 & -2.78E+01 &  1.89E+00 & -9.66E-02 &  1.04E+01 & 1.2E-03\\
cf  274   98 & 1 & -4.50E+01 &  0.00E+00 &  0.00E+00 &  6.69E+01 & -8.82E+00 &  6.52E-01 & -8.66E+00 & 1.5E-03\\
cf  274   98 & 0 &  3.45E+01 &  0.00E+00 &  0.00E+00 & -3.97E+01 &  3.13E+01 & -1.60E+01 &  3.53E+00 & 1.5E-03\\
cf  275   98 & 0 &  8.34E+00 &  1.13E-02 & -2.48E+00 &  1.50E+01 & -1.35E+00 &  8.30E-02 & -3.67E+00 & 1.1E-03\\
cf  276   98 & 1 & -3.74E+01 &  0.00E+00 &  0.00E+00 &  5.39E+01 & -8.74E+00 &  6.38E-01 &  0.00E+00 & 1.5E-02\\
cf  276   98 & 0 &  8.18E+00 &  0.00E+00 &  0.00E+00 & -2.88E+00 &  6.77E+00 & -2.68E+00 &  0.00E+00 & 1.5E-02\\
cf  277   98 & 1 &  8.74E+01 &  0.00E+00 &  0.00E+00 & -7.63E+01 &  3.03E+00 & -7.28E-02 &  3.10E+01 & 1.2E-03\\
cf  277   98 & 0 &  1.87E+01 &  0.00E+00 &  0.00E+00 & -4.86E+00 &  4.23E+00 & -1.66E+00 &  4.15E-01 & 1.2E-03\\
\hline\hline
\end{tabular}
\end{table*}

This section is intended to help with interpreting the information given in the
electronic tables. We also give examples for calculating the reaction rate for
a given reaction and its inverse reaction.

Table \ref{Tab_a1}    shows which  information is contained in the rate
Tables 3--6 for different mass predictions, available at the CDS (Table
\ref{Tab_a1} illustrated some lines from CDS Table 3:  rates on the basis of
ETFSI-predictions).
 The organization of Tables  3--6 with calculated rates for 24 different
values of $T_9$ is extremely simple -- all data are in   eight  columns. The columns
give, in this order, the target element, the atomic mass number $A$ of the
target, the temperature in $T_9$,  the  partition function of the target,
  the  neutron-induced fission rate for the ground state,
  the  neutron-induced fission rate
with thermally populated target states,  the  neutron capture rate for the ground
state,  and the  neutron capture rate with thermally populated target states (units
of all rates: $cm^{-3}mole^{-1}c^{-1}$).

The rates in Tables  3--6 were calculated on the basis of different mass and
fission barrier predictions: ETFSI, TF, FRDM(masses)+TF(barriers) and HFB
respectively (see main text of the paper for details).

 The entries
are denoted as follows:

\begin{lyxlist}{00.00.0000}
\item [mother] mother nucleus (target),
\item [$T_9$] stellar temperature in 10$^9$ K,
\item [p.f.] nuclear partition function,
\item [(n,g)](n,\( \gamma  \))-rate, $N_{A}\left\langle \sigma_{n\gamma}  v\right\rangle$
\item [(n,g)*](n,\( \gamma  \))-rate, $N_{A}\left\langle \sigma_{n\gamma}  v\right\rangle$, with thermally populated
target levels,
\item [(n,f)]   (n,fission)-rate, $N_{A}\left\langle \sigma_{nf}  v\right\rangle$
\item [(n,f)*]  (n,fission)-rate, $N_{A}\left\langle \sigma_{nf}  v\right\rangle$, with thermally populated
target levels.
\end{lyxlist}

 Table \ref{Tab_a2}   shows  which information is contained in the fit
Tables 7--18,   available at the CDS.  The following information is provided:
\begin{lyxlist}{00.00...0000}
\item [mother]reaction target
\item [Dev]fit accuracy \( \zeta  \) (Eq.\ (\ref{eq:accur}))
\item [\( a_{0} \)\ldots{}\( a_{6} \)]seven fit parameters for the
forward rate
\item [\( a_{0}^{\mathrm{rev}}\)\ldots{}\(a_{6}^{\mathrm{rev}}\)]
seven fit parameters for the reverse rate fit (see Sect.~\ref{sec:revpar})
\end{lyxlist}

The fitting coefficients for the (neutron,$\gamma$)-, ($\gamma$,n)-, and
neutron-induced fission rates with different mass and fission-barrier
predictions are placed in the Tables 7--18.
The columns in Tables 7--18 are organized as follows: target element, atomic
mass number $A$ of the target, target charge number $Z$, the number of fitting
curves $i_{fit}$, seven coefficients of the forward reaction $a_i$, and the
mean square error.

A value $i_{fit}$=0 means that there is only one seven-parameter set to fit the
rate. Values $i_{fit}$>1 give the number of parameter sets which have to be
added up to yield the final rate, i.e.\ the rate $r$ is calculated as $r=\sum_i
r_i$, with each $r_i$ computed from the $i$-th parameter set and using Eq.\
(\ref{eq_fitpar}).

 The examples of Tables  3--6  are given for 1 isotope for 24 values of
$T_9$, and for Tables 7--18 - for 10 isotopes. Note that the tables of CDS show
the values with an accuracy of  seven  digits.

Below we give two examples for calculating a rate at $T_9=7.0$ with the fit
parameters listed in the tables.

  The first example is the reaction \( ^{273} \)Cf(n,\( f \)). In
Table A.2 one finds the parameters \( a_{0}= 38.72 \), \( a_{1}=-0.057 \), \(
a_{2}=8.78 \), \( a_{3}=-27.85 \), \( a_{4}= 1.90 \), \( a_{5}=-0.097 \), \(
a_{6}= 10.45 \). With the help of Eq.\ (\ref{eq_fitpar}) one calculates \(
N_{A}\left\langle \sigma v\right\rangle ^{*}_{fit}=1.574 \times 10^9$ cm\( ^{3}
\)s\( ^{-1}\)mole\(^{-1} \) or \(\mathrm{Log}_{10}( N_{A}\left\langle \sigma
v\right\rangle ^{*}_{fit}) =9.20 \) at \( T_{9}=7.0 \).

The second example is for the rates of the capture reaction\( ^{273} \)Cf(n,\(
\gamma \))\( ^{274} \)Cf  and its reverse reaction. Similar to the above
example, using the parameters from Table A.2 and Eq.\ (\ref{eq_fitpar}) the
capture rate is easily found to be
 \( N_{A}\left\langle \sigma
v\right\rangle ^{*}_{fit}=120\) cm\( ^{3} \)s\( ^{-1}\)mole\(^{-1} \).    With
the reverse parameters the first value in the determination of the reverse rate
is found to be \( \lambda_\gamma \)\( ^{\prime }= \)5.0\( \times \)10\( ^{9} \)
s\( ^{-1} \) at \( T_{9}=7.0 \). In order to obtain the actual value of the
reverse rate, one first has to determine the ratio of the partition functions
$G_{^{273}\mathrm{Cf}} / G_{^{274}\mathrm{Cf}} = 4.58\times 10^5 / 3.87\times
10^5=1.18$ (see Table A.1 and Sec.~\ref{sec:revpar}). The value
$\lambda^\prime$ has to be multiplied by this ratio to derive the
photodisintegration rate $\lambda$:

\[
\lambda_\gamma =\lambda_\gamma ^{\prime }\frac{G_{^{273}\mathrm{Cf}}}{
G_{^{274}\mathrm{Cf}}}=\lambda_\gamma ^{\prime } \times 1.18=5.9\times
10^{9}\quad \mathrm{s}^{-1}.\] The values of the partition functions at \(
T_{9}  \) are also given in the
 online Tables 3--6.
Note that the procedure is always the same as described above, regardless of
whether it is an exoergic or an endoergic reaction.

\end{appendix}

\bibliographystyle{aa} 
\bibliography{11967main}

\end{document}